\documentclass{article}

\usepackage{amsmath}
\usepackage{amssymb}
\usepackage{amsthm}
\usepackage{color}
\usepackage{graphicx}

\usepackage[english]{babel}

\begin{document}

\title{Entanglement in dissipative dynamics of identical particles}

\author{U. Marzolino\\
\\
\small{Albert-Ludwigs-Universit\"at Freiburg, Hermann-Herder-Stra\ss e 3, D-79104 Freiburg, Germany} \\
\small{Univerza v Ljubljani, Jadranska 19, SI-1000 Ljubljana, Slovenija}
}

\date{}

\maketitle

\begin{abstract}
Entanglement of identical massive particles recently gained attention, because of its relevance in highly controllable systems, \emph{e.g.} ultracold gases. It accounts for correlations among modes instead of particles, providing a different paradigm for quantum information. We prove that the entanglement of almost all states rarely vanishes in the presence of noise, and analyse the most relevant noise in ultracold gases: dephasing and particle losses. Furthermore, when the particle number increases, the entanglement decay can turn from exponential into algebraic.
\end{abstract}

Quantum correlations were proved to be a key resource in quantum information processing \cite{NielsenChuang,Horodecci,Buhrman2010} and a useful tool for studying condensed matter systems \cite{Lewenstein2007,Amico2008}. Many of these studies were developed in the framework of distinguishable particles, where particles are manipulated locally. The peculiarity of identical particles is that they cannot be individually addressed, unless particles are effectively distinguished employing additional degrees of freedom \cite{Herbut2001,Herbut2006,Tichy2013}, \emph{e.g.} confining them in different positions.

The behaviour of truly identical particles is of practical importance, since they are the elementary constituents of several physical systems in atomic and condensed matter physics. For instance, ultracold gases \cite{Leibfried2003,Bloch2008,Giorgini2008,Negretti2011} can be controlled with a very high precision, and are a promising arena for the study of many-body physics and applications in quantum information \cite{Ionicioiu2002,Giorda2004,Anders2006,Benatti2010,Benatti2011,Argentieri2011}. However, these systems are unavoidably affected by noise, typically dephasing and particle losses \cite{Leibfried2003}.

Despite different approaches \cite{Schliemann2001,Ghirardi2002,Eckert2002,Wiseman2003,Tichy2013}, only a few of them are relevant for observable predictions. The characterization of quantum correlations with identical particles cannot rely on the tensor product structure of single particle Hilbert spaces, but must be reformulated in terms of subsets of locally manipulable observables \cite{Zanardi2001,Zanardi2002-1,Barnum2004,Narnhofer2004,Benatti2010,Benatti2012,Benatti2012-2}. This powerful approach generalizes the theory of entanglement of distinguishable particles. Moreover, when applied to identical particles, it accounts for quantum correlations of occupations of orthogonal modes, like wells in optical lattices or hyperfine levels of molecules, which are individually addressable in actual experiments \cite{Wurtz2009,Gross2010,Riedel2010}.

Within this framework, we shall prove that entanglement of almost all the states is never completely dissipated by noise under minimal assumptions including any Markovian noise, contrary to the case of distinguishable particles. Moreover, entanglement is easily generated by tunneling even in the presence of noise. We shall show with concrete examples that entanglement can decay exponentially in time but is never lost. Interestingly, when the number of particles is very large the exponential decay can turn into an algebraic decay.

Beyond the characterization of the dynamics itself, the non-vanishing entanglement of many replicas of states discussed here, though small, can be distilled into a fewer copies of more entagled states via local operations and classical communiation, which are useful for the aforementioned applications (\emph{e.g.} see refs. \cite{Schuch2004,Sangouard2008,Gonta2012,Sheng2013}). The present results pave the way for the generalization of entanglement dissipation \cite{Zyczkowski2001,Benatti2003,Carvalho2004,
Almeida2007,Benatti2008,Yu2009} and the characterization of entanglement distroying noise \cite{Horodecki2003,Holevo2008,Filippov2013} to identical particles.

\section{Entanglement}

We deal with many-body systems, whose constituents are $N$ bosons which can fill $M$ orthogonal modes. In the more convenient formalism of second quantization, $a^\dagger_j,a_j,\,j=1,2,\ldots,M$ are the creation and annihilation operators for each mode, satisfying the commutation relations $[a_j,\,a^\dagger_l]=\delta_{jl}$. The total Hilbert space $\cal H$ of the system is spanned by the Fock states

\begin{equation} \label{fock.states}
|k_1, k_2,\cdots,k_M\rangle= \frac{(a_1^\dagger)^{k_1}\,(a_2^\dagger)^{k_2}\,\cdots\,(a_M^\dagger)^{k_M}\,|0\rangle}{\sqrt{k_1!\,k_2!\cdots k_M!}},
\end{equation}
where $k_j$ is the occupation number of the $j$-th mode and $\sum_{j=1}^M k_j=N$. The norm-closure of the set of polynomials in all creation and annihilation operators is the algebra of all bounded operators ${\cal B}({\cal H})$, including all the observables. We define an \emph{algebraic bipartition} of this algebra by splitting the set of creation and annihilation operators into two disjoint sets
$\{a_j^\dagger,a_j\,|\,j=1,2,\dots,m\}$ and 
$\{a_j^\dagger,a_j\,|\,j=m+1,m+2,\dots,M\}$. The norm-closures of all polynomials in each set are two commuting subalgebras, ${\cal A}_1, {\cal A}_2\subset {\cal B}({\cal H})$: any
element of ${\cal A}_1$ commutes with any element of ${\cal A}_2$, $[{\cal A}_1, {\cal A}_2]=0$. This is the fundamental property of locality: measurements on different subalgebras do not influence each other. An operator is said to be \emph{local} with respect to the bipartition $({\cal A}_1, {\cal A}_2)$, if it is the product $A_1 A_2$ of an operator 
$A_1$ in ${\cal A}_1$ and another $A_2$ in ${\cal A}_2$. A pure state $|\psi\rangle$ is said to be \emph{separable} with respect to the bipartition $({\cal A}_1,{\cal A}_2)$ if

\begin{equation}
|\psi\rangle={\cal P}(a^\dagger_1, \dots, a^\dagger_m)\cdot 
{\cal Q}(a^\dagger_{m+1},\ldots ,a^\dagger_M)\ |0\rangle, \label{sep}
\end{equation}
where ${\cal P}$ and ${\cal Q}$ are arbitrary polynomials. Mixed separable states are convex combinations of pure separable states. Otherwise, the state is \emph{entangled}. See \cite{Benatti2010,Benatti2012} for a detailed analysis.

In our framework, the local subsystems are groups of modes rather than particles. Unlike modes, identical particles are not individually addressable, and commuting subalgebras of physical observables on different particles do not exist. Thus, our definition of entanglement is based on the operational identification of individually measurable portions of the physical system. At variance, some of the other approaches rely only on the formal factorizability in first quantization and disregard the physical implications of indistinguishability.

Since the total number of massive particles is conserved in atomic and condensed matter systems, all the observables and the density matrices commute with the total number operator. There is a superselection rule \cite{Bartlett2007}, which forbids any coherent superpositions of states with different total numbers of particles. This is crucial in the characterizations of entanglement, based on the partial transposition criterion and the robustness against mixtures, providing a new geometrical picture of entangled states \cite{Benatti2012,Benatti2012-2}. Any orthonormal basis of separable states can be relabelled as

\begin{equation}
|k,\sigma; N-k,\sigma'\rangle, \quad \sigma=1,2, \dots , {k+m-1 \choose k}, \quad \sigma'=1, 2, \dots , {N-k+M-m-1 \choose N-k}. \label{sep.basis}
\end{equation}
The integer $k$ counts the number of particles in the first $m$ modes,
while $\sigma$ labels the different ways in which $k$ particles can fill those modes. Similarly, $\sigma'$ labels the ways in which the remaining $N-k$ particles can occupy the other $M-m$ modes. Any state can be expanded in such basis:

\begin{equation} \label{state}
\rho=\sum_{k,l=0}^N\sum_{\sigma,\sigma',\tau,\tau'}\rho_{k \sigma\sigma', l\tau\tau'}\ |k,\sigma;N-k,\sigma'\rangle\langle l,\tau;N-l,\tau'|.
\end{equation}

Previous works \cite{Benatti2012,Benatti2012-2} pointed out two qualitatively different contributions to entanglement. The first is the coherence among different labels $\sigma,\sigma'$, keeping $k$ fixed, which is qualitatively and mathematically analogous to the entanglement of distinguishable particles \cite{Benatti2012,Benatti2012-2}. Physically, the local number of particles is fixed in such states, thus the mode-bipartition corresponds to a particle-bipartition. A state which exhibits only these coherences has a block-diagonal structure in the label $k$: $\rho_{k\sigma\sigma',l\tau\tau'}=0$ if $k\neq l$. The set of block-diagonal states has zero measure and includes all the separable and positive under partial transposition (PPT) states \cite{Benatti2012}. The second contribution comes from the coherence among different $k$. In this case, the particle-bipartition is not physically accessible. These states are not block-diagonal and thus not PPT \cite{Benatti2012}.

Experimental advances in controlling this latter kind of entanglement has been developped with ultracold atoms \cite{Gross2010,Riedel2010}. Identical particles have already been proposed for implementing quantum information processing \cite{Ionicioiu2002,Negretti2011}. However, these proposals are based only on the first contribution of entanglement, \emph{i.e.} states formally equivalent to those of distinguishable particle are encoded in the restricted class of block-diagonal states.

In this paper, we discuss the entanglement of non-block-diagonal states under dissipative dynamics \cite{BreuerPetruccione,Benatti2005}. We shall estimate entanglement via negativity ${\cal N}(\rho)$, namely the entanglement measure derived from the partial trasposition criterion \cite{Vidal2002}. The negativity of the state \eqref{state} was computed in \cite{Benatti2012}:

\begin{eqnarray}
\label{neg}
{\cal N}(\rho) & = & \frac{1}{2}\left(\sum_{k,l=0}^N{\rm Tr}\big(\sqrt{{\mathcal R}_{k,l}}\big)-1\right), \\
{\mathcal R}_{k,l} & = & \sum_{\substack{\sigma,\sigma',\sigma'',\\ \tau,\tau',\tau''}}\rho_{k\sigma''\sigma,l\tau\tau''} \, \overline{\rho_{k\sigma'\sigma,l\tau\tau'}}|k,\sigma';N-l,\tau'\rangle\langle k,\sigma'';N-l,\tau''|,
\end{eqnarray}
where the bar is the complex conjugate. The entanglement of mixtures of different particle numbers $P=\sum_N p_N\rho^{(N)}$ is quantified by the average ${\cal N}(P)=\sum_N p_N{\cal N}(\rho^{(N)})$ \cite{Benatti2012}.

\section{General noise}

We study the effects of undesired interactions with the environment resulting in non-unitary time-evolutions \cite{BreuerPetruccione,Benatti2005}. The following property on entanglement dynamics holds for general time-evolutions under minimal assumptions.

{\bf Proposition.} \emph{If the dynamical map $\rho_0\to\rho_t$ cannot superimpose different blocks and is one-to-one for any finite $t$ then the entanglement of non-block-diagonal states can vanish only asymptotically in time. \\
\indent
If the dynamics can superimpose different blocks and its time derivative $\partial\rho_t/\partial t$ is continuous, then the entanglement of non-block-diagonal states vanishes at most at discrete times.}

\proof If different blocks cannot be superimposed, the diagonal blocks evolve independently from the off-diagonal ones. If the time-evolution of a non-block-diagonal state were disentangled at a certain time, it would be block-diagonal. It would be the same time-evolution of a different initial state made only of the diagonal blocks of the actual initial state. This is in contradiction with the assumptions.

In the second case, consider the first time $t_0$ when entanglement vanishes. At the first order in the infinitesimal time interval $\delta$, $\rho_{t_0}=\rho_{t_0-\delta}+\delta\left.\partial\rho_t/\partial t\right|_{t\to t_0}$ is separable, hence, block-diagonal. Thus, the off-diagonal blocks of $\delta\left.\partial\rho_t/\partial t\right|_{t\to t_0}$ are not zero and cancel those of $\rho_{t_0-\delta}$. Finally, at the same order $\rho_{t_0+\delta}=\rho_{t_0}+\delta\left.\partial\rho_t/\partial t\right|_{t\to t_0}$ is non-block-diagonal, hence, entangled. \hfill $\blacksquare$

An immediate consequence is that non-block-diagonal states are generated out of separable states if and only if the dynamics induces tunneling between the groups of modes that define the subsystems. In other words, tunneling and entanglement in the form of off-diagonal blocks are equivalent.

We now consider the most studied non-unitary time-evolutions, namely those described by Lindblad master equations \cite{BreuerPetruccione,Benatti2005}: $d\rho_t/dt=L[\rho_t]$ with

\begin{equation} \label{noise}
L[\rho_t]=-\frac{i}{\hbar}[H,\rho_t]+\sum_j \lambda_j\left(A_j\rho_t A_j^\dag-\frac{1}{2}\left\{A_j^\dag A_j,\rho_t\right\}\right),
\end{equation}
where $H$ is the hamiltonian of the system, and $A_j$ are the so-called Lindblad operators. This master equation describes a markovian dynamics, whose underlying interaction with the environment depends on the operators $A_j$ and the positive coupling constants $\lambda_j\geqslant0$. The conservation of the total number of particles imposes limitations to the Lindblad operators, \emph{i.e.} $A_j$ cannot superimpose states with different total numbers of particles. The previous Proposition applies to the solution of eq. \eqref{noise}: $\rho_t=e^{tL}[\rho]$. More general master equations are not always one-to-one maps, although there are sufficient conditions for non-linear mean-field \cite{Alicki1983} and time-dependent non-Markovian \cite{Chruscinski2010} generators. See \cite{Matisse2008} for a concrete exact non-Markovian dynamics, and \cite{Liu2011} for an experimental realization. The above relation between tunneling and entanglement suggests that general time-evolved entanglement is detected by particle fluctions or currents \cite{Li2005,Welack2006,Francis2012}.

It is worthwhile to note differences with distinguihable particles. In the latter case, dissipations that locally affect each subsystem can disentangle the state at finite time, which cannot be re-entangled \cite{Zyczkowski2001,Carvalho2004,Almeida2007,Yu2009}. Moreover, if the interaction between subsystems induced either by the hamiltonian or by the Lindblad operators is much weaker than the local noise, the entanglement cannot be generated \cite{Benatti2003,Benatti2008}. These two phenomena do not occur here.

We now exemplify the entanglement dissipation induced by the two main sources of noise in ultracold atomic gases: particle losses and dephasing \cite{Leibfried2003}. Although the time-evolution of these noises is not analytically known for general hamiltonians, we derive information on the entanglement dynamics independently from the hamiltonian. We use the Trotter formula

\begin{equation}
e^{A+B}=\lim_{n\to\infty}\left(e^{\frac{A}{n}}e^{\frac{B}{n}}\right)^n,
\end{equation}
where $A,B$ be operations on density matrices.

\section{Particle losses}

Particle losses are modelled by Lindblad operators that are polynomials of the annihilation operators \cite{Li2009}. By expanding $\rho_t=e^{tL}[\rho]$ in Taylor series and collecting the contributions, the time-evolution is

\begin{eqnarray}
\label{sol.loss} & & \rho_t=e^{tL}[\rho]=e^{t\left(L_{\rm ham}+L_{\rm damp}\right)}[\rho]+\sum_{N'=0}^{N-1}\rho_t^{(N')}, \\
& & L_{\rm ham}[\rho]=-\frac{i}{\hbar}[H,\rho], \\
& & L_{\rm damp}[\rho]=-\sum_j\frac{\lambda_j}{2}\{A_j^\dag A_j,\rho\},
\end{eqnarray}
where \eqref{state} is the inital state, $e^{t\left(L_{\rm ham}+L_{\rm damp}\right)}[\rho]$ is the contribution from the sector with $N$ particles, and $\rho_t^{(N')}$ are matrices lying on the subspace of $N'<N$ particles and resulting from the terms $A_j\rho_tA_j^\dag$ of the generator. Consider a hamiltonian with block-diagonal eigenvectors and local operators $A_j$, \emph{e.g.} $a_j^m$ or $a_j a_l$ as discussed in \cite{Li2009}. This dynamics can disentangle a block-diagonal entangled state at finite times, as shown with the following example.

In the case of four equally bipartite modes ($M=2m=4$) with only one operator $A_0=a_1a_3$ and hamiltonian $H=\sum_{j=1}^4 \varepsilon_ja_j^\dag a_j$. The initial state

\begin{equation} \label{example}
\rho=p \, |\psi_1\rangle\langle\psi_1|+\frac{1-p}{2}\left(|\psi_2\rangle\langle\psi_2|+|\psi_3\rangle\langle\psi_3|\right),
\end{equation}
with $0\leq p\leq1$, and

\begin{equation}
|\psi_1\rangle=\frac{1}{\sqrt{2}}\big(|0,1,0,1\rangle+|1,0,1,0\rangle\big), \qquad |\psi_2\rangle=|0,1,1,0\rangle, \qquad |\psi_3\rangle=|1,0,0,1\rangle,
\end{equation}
has support only on the block $\rho^{(1,1)}$ of one particle in each group of two modes, thus it is block-diagonal. Applying the partial transposition criterion \cite{Benatti2012}, it is entangled if and only if $p>\frac{1}{2}$. The evolved state

\begin{equation}
\gamma_t[\rho]=\frac{p}{2}\left(1-e^{-t\lambda_0}\right)|0\rangle\langle 0|+p \, |\psi_1(t)\rangle\langle\psi_1(t)|+\frac{1-p}{2}\big(|\psi_2\rangle\langle\psi_2|+|\psi_3\rangle\langle\psi_3|\big),
\end{equation}
with the non-normalized state

\begin{equation}
|\psi_1(t)\rangle=\frac{1}{\sqrt{2}}\Big(e^{-\frac{it}{\hbar}(\varepsilon_2+\varepsilon_4)}|0,1,0,1\rangle+e^{-t\left(\frac{\lambda_0}{2}+\frac{i}{\hbar}(\varepsilon_1+\varepsilon_3)\right)}|1,0,1,0\rangle\Big),
\end{equation}
is separable if and only if $e^{-t\lambda_0/2}p\leq 1-p$, according to the partial transposition criterion \cite{Benatti2012}. The initial entanglement is completely dissipated for sufficiently large $t$.

Consider now the time-evolution of entanglement of non-block-diagonal states with local operators $A_j$. We focus on the contribution of the sector with $N$ particles in \eqref{sol.loss}. The time-evolved entanglement is bounded from below by its hamiltonian dynamics alone, damped by an exponential factor. Since the exponential factor vanishes only at infinite times in the presence of non-vanishing noise, the time-evolution of entanglement qualitatively traces the hamiltonian evolution. Thus, the noise itself is not able to erase entanglement.

In order to prove this statement, apply the Trotter formula with $A=tL_{\rm ham}$ and $B=tL_{\rm damp}$ in \eqref{sol.loss}. Using $|(e^{\frac{t}{n}L_{\rm damp}}[\rho])_{k\sigma\sigma',l\tau\tau'}|\geqslant e^{-\frac{t}{n}\eta}|\rho_{k\sigma\sigma',l\tau\tau'}|$ with $\eta$ the maximum eigenvalue of $\sum_j\lambda_j A_j^\dag A_j$ whose eigenspace belongs to the sector of $N$ total particles, we get

\begin{equation} \label{bound.loss}
{\cal N}\left(\rho_t\right)={\cal N}\left(e^{t\left(L_{\rm ham}+L_{\rm damp}\right)}[\rho]\right)+\sum_{N'=0}^{N-1}{\cal N}\left(\rho_t^{(N')}\right)\geqslant e^{-t\eta}{\cal N}\left(e^{tL_{\rm ham}}[\rho]\right).
\end{equation}

If all the modes are independently affected by particle losses, \emph{e.g.} for $A_j=a_j$, there is only one stationary state, namely the vacuum $|0\rangle\langle 0|$ which is separable. A stationary state is also an asymptotically stable state. The theorem proved in \cite{Schirmer2010} states that if there is a unique stationary state, then it is also the unique asymptotic state. Thus, the asymptotic state is separable.

\section{Dephasing}

We now focus on dephasing \cite{Palma1996,Ferrini2010,Argentieri2011}, where the Lindblad operators are the local number operators $A_j=a_j^\dag a_j$, and the coherences between the Fock states are damped. A realization of such noise is the fluctuation of the depth of the wells, representing the modes, in an optical lattice. Consider that $H$ has block-diagonal eigenvectors. This dynamics can disentangle a block-diagonal state at finite times, as shown via the following example.

Consider the case of $H=\sum_{j=1}^4 \varepsilon_j a_j^\dag a_j$, the time-evolution of the entangled state \eqref{example}, namely

\begin{eqnarray}
\gamma_t[\rho] & = & \frac{1-p}{2}\Big(|\psi_2\rangle\langle\psi_2|+|\psi_3\rangle\langle\psi_3|\Big)+\frac{p}{2}e^{-t\sum_{j=1}^4 \frac{\lambda_j}{2}}\Big(e^{\frac{i}{\hbar}t(\varepsilon_2+\varepsilon_4-\varepsilon_1-\varepsilon_3)}|1,0,1,0\rangle\langle 0,1,0,1|+{\rm h.c.}\Big) \nonumber \\
&& +\frac{p}{2}\Big(|0,1,0,1\rangle\langle 0,1,0,1|+|1,0,1,0\rangle\langle 1,0,1,0|\Big), \nonumber \\
\end{eqnarray}
is separable if and only if $e^{-t\sum_{j=1}^4 \lambda_j/2}p\leq(1-p)$, according to the partial transposition criterion \cite{Benatti2012}. Thus, the initial entanglement is completely erased for sufficiently large $t$, in analogy with distinguishable particles \cite{Zyczkowski2001,Carvalho2004,Almeida2007,Yu2009}.

As for particle losses, the time-evolved entanglement of an arbitrary non-block-diagonal is bounded by an exponential damping of its hamiltonian dynamics. However, dephasing alone is not able to disentangle non-block-diagonal states in finite times. The Trotter formula applied to $e^{tL}$ with $A=tL_{\rm ham}$ and $B=t(L-L_{\rm ham})$, and the equalities

\begin{equation}
\left|\langle k_1,\dots,k_M|e^{t(L-L_{\rm ham})}[\rho]|\bar k_1,\dots,\bar k_M\rangle\right|=e^{-t\sum_{j=1}^M\frac{\lambda_j}{2}(k_j-\bar k_j)^2}\left|\langle k_1,\dots,k_M|\rho|\bar k_1,\dots,\bar k_M\rangle\right| \nonumber
\label{decoherence}
\end{equation}
imply

\begin{equation} \label{bound.deph}
{\cal N}(\rho_t)\geqslant e^{-t N^2\sum_{j=1}^M\frac{\lambda_j}{2}}{\cal N}\left(e^{tL_{\rm ham}}[\rho]\right).
\end{equation}

According to a Frigerio's theorem \cite{Frigerio1978}, if there is a full rank stationary state and all the operators which commute with $H$ and with all $\lambda_j a_j^\dag a_j$ are proportional to the identity, then the stationary state is unique. Moreover, if a Lindblad master equation has a unique stationary state, it is also the unique asymptotic state for every initial state \cite{Schirmer2010}. Let us now apply the Frigerio's theorem to show sufficient conditions in order the asymptotic state to be separable. Consider an initial state with $N$ particles, and the set ${\cal M}_0=\{j\,|\,\lambda_j=0\}$ of modes which do not feel the dephasing. We write an arbitrary operator as the sum of its hermitian and anti-hermitian parts: $O=O_H+O_A$, where $O_H=\frac{O+O^\dag}{2}$ and $O_A=\frac{O-O^\dag}{2}$. The condition $[a_j^\dag a_j,O]=[a_j^\dag a_j,O_H]+[a_j^\dag a_j,O_A]=0$ is satisfied if and only if $[a_j^\dag a_j,O_H]=[a_j^\dag a_j,O_A]=0$, because $[a_j^\dag a_j,O_H]$ is an anti-hermitian operator and $[a_j^\dag a_j,O_A]$ is hermitian. This implies that both $O_H$ and $O_A$ must have common eigenvectors with each $a_j^\dag a_j$ such that $j\notin{\cal M}_0$, namely the Fock states in the modes $j\notin{\cal M}_0$ factorized with arbitrary states in the modes $j\in{\cal M}_0$. Applying the same argument to $[H,O]=0$, $O$ has to be diagonal also in the eigenbasis of $H$. If subgroups of modes cannot be factorized in the eigenstates of $H$, \emph{e.g.} if $H=-\sum_{j=1}^{M-1} \tau_j(a_j^\dag a_{j+1}+a_j a_{j+1}^\dag)$, the only operators which commute with $H$ and all $\lambda_j a_j^\dag a_j$ are proportional to the identity operator. The unique asymptotic state is the identity operator ${\bf 1}_N$ with support on the subspace of $N$ particles, since $L_{\rm deph}[{\bf 1}_N]=0$. In the most general case, the initial state is a mixture of states with different number of particles $\sum_N p_N\rho^{(N)}$, with $p_N\geq 0$, $\sum_N p_N=1$, $\rho^{(N)}\geq 0$ and ${\rm Tr}\,\rho^{(N)}=1$. Then, the asymptotic state is $\sum_N p_N{\bf 1}_N$, which is separable. Notice that if $\lambda_j\neq 0$ for at least $M-1$ modes the asymptotic state is separable even if the hamiltonian has the same eigenvalues of $a_j^\dag a_j$. In this case the dynamics is a mere decoherence in the Fock basis, and the asymptotic state is a mixture of Fock states, thus separable.

\section{Large $N$}

In the previous examples, negativity has different contibutions
which exponentially decay with several rates, and the higher rate gives the lower bounds \eqref{bound.loss} and \eqref{bound.deph}. If the total number of particles $N$ is large, occupation numbers are approximated by continuous variables, and their sums by integrals. As a result, the decay rates are very close to each other and the overall decay can be algebraic. This change substantially delays the onset of the regime with infinitesimally small entanglement.

We show this phenomenon class of states defined by $\rho_{k\sigma\sigma',l\tau\tau'}=\rho_{k\sigma\sigma',l\sigma\sigma'}\delta_{\sigma,\tau}\delta_{\sigma',\tau'}$ in eq. \eqref{state}, affected by dephasing and hamiltonian $H=\sum_{j=1}^M\varepsilon_ja_j^\dag a_j$. This class of states is preserved by the dynamics. Furthermore, consider a subset of the Fock basis $(k,\sigma,\sigma')=\left\{k_j\right\}_j$ and $(l,\sigma,\sigma')=\left\{l_j\right\}_j$, where either $k_j=l_j+c_j|k-l|^\alpha$ if $l_j\leqslant k_j$ or $l_j=k_j+c_j|k-l|^\alpha$ if $k_j\leqslant l_j$. The non-negative coefficients $c_j\geqslant0$ are determined by $\sum_{j=1}^mk_j=k$, $\sum_{j=1}^ml_j=l$, and $\sum_{j=1}^Mk_j=\sum_{j=1}^Ml_j=N$. The time-evolution is

\begin{equation}
(\rho_t)_{k\sigma\sigma',l\sigma\sigma'}=e^{-i\frac{t}{\hslash}\sum_{j=1}^M\varepsilon_j(k_j-l_j)-\frac{t}{2}\sum_{j=1}^M\lambda_j(k_j-l_j)^2}\rho_{k\sigma\sigma',l\sigma\sigma'},
\end{equation}
and the negativity of the evolved state is

\begin{equation}
{\cal N}(\rho_t)=\frac{1}{2}\left(\sum_{k,l}\sum_{\sigma,\sigma'}\left|(\rho_t)_{k\sigma\sigma',l\sigma\sigma'}\right|-1\right).
\end{equation}
Now, define the variables $x=\frac{k+l}{2N}$ and $y=\frac{k-l}{N}$, that range in the interval $[0,1]$ and tend to continuous variables in the limit of large $N$. Define also the function $R_{\sigma\sigma',\sigma\sigma'}(x+y/2,x-y/2)=N\rho_{k\sigma\sigma',l\sigma\sigma'}$, with the normalization $1=\sum_{k=0}^N\sum_{\sigma,\sigma'}\rho_{k\sigma\sigma',k\sigma\sigma'}\simeq\int_0^1 dx\sum_{\sigma,\sigma'}R_{\sigma\sigma',\sigma\sigma'}(x,x)$. The sums over $k$ and $l$ is then well approximated by integrals, and with $S=\sum_{j=1}^M\lambda_jc_j$, we get

\begin{equation} \label{neg.approx}
{\cal N}(\rho_t)\simeq\frac{1}{2}\Bigg(N\int_0^1 dx\int_{2\left|x-\frac{1}{2}\right|-1}^{1-2\left|x-\frac{1}{2}\right|} dy \, e^{-\frac{t}{2}N^{2\alpha}Sy^{2\alpha}}\sum_{\sigma,\sigma'}\left|R_{\sigma\sigma',\sigma\sigma'}\left(x+\frac{y}{2},x-\frac{y}{2}\right)\right|-1\Bigg).
\end{equation}

For large $tSN^{2\alpha}\gg1$, the exponential function in the integrand is highly peaked around $y=0$ and asymptotic expansions of integrals \cite{Wong} can be applied. If the dependence of the rest of the integrand from $N$ is subleading with respect to the exponential, we can further approximate the negativity by substituting the domain of the integration in $dy$ with $(-\infty,\infty)$. We now expand the function $R$ in Taylor series around the peak of the exponential function in the integrand:

\begin{equation}
R_{\sigma\sigma',\sigma\sigma'}\left(x+\frac{y}{2},x-\frac{y}{2}\right)=\sum_{n=0}^\infty\frac{y^n}{n!}\left.\frac{\partial^n}{\partial y^n}R_{\sigma\sigma',\sigma\sigma'}\left(x+\frac{y}{2},x-\frac{y}{2}\right)\right|_{y=0}.
\end{equation}
Plugging this expansion in eq. \eqref{neg.approx}, the integration in $dy$ can be performed. Using the normalization, we get the final result:

\begin{equation} \label{neg.approx2}
{\cal N}(\rho_t)\simeq -\frac{1}{2}+\frac{2^{\frac{1}{2\alpha}}\Gamma\left(1+\frac{1}{2\alpha}\right)}{(tS)^{\frac{1}{2\alpha}}}+\sum_{n=1}^\infty\frac{2^{\frac{2n+1}{2\alpha}}\Gamma\left(1+\frac{2n+1}{2\alpha}\right)}{(2n+1)!N^{2n}(tS)^{\frac{2n+1}{2\alpha}}}\sum_{\sigma,\sigma'}\int_0^1 dx\left[\left|\frac{\partial^{2n}}{\partial y^{2n}}R_{\sigma\sigma',\sigma\sigma'}\left(x+\frac{y}{2},x-\frac{y}{2}\right)\right|\right]_{y=0},
\end{equation}
where $\Gamma(\cdot)$ is the gamma function.

For large $N$, the negativity is approximated by a truncation of the series in \eqref{neg.approx2}, and decays algebraically in time. The above condition $tSN^{2\alpha}\gg1$ implies that the leading order of \eqref{neg.approx2} is $\ll N2^{\frac{1}{2\alpha}}\Gamma(1+\frac{1}{2\alpha})$, compatible with any scaling $N^p$ and $0<p<1$ which increases with the particle number. Notice that, if $tS\gg1$ the width of the exponential in the integrand $\sim\frac{1}{N(tS)^{1/(2\alpha)}}$ is much smaller than the discrete spacing of the variable $y$, that is $1/N$. This forbids us to approximate the sums with integrals.

While the above class of states has zero measure for $1<m<M-1$, it includes any state with $m=1$ or $m=M-1$. For two-mode particles ($M=2m=2$), \emph{e.g.}, the only choice is $\alpha=c_{1,2}=1$, and 
the estimation \eqref{neg.approx2} holds for any two-mode state such that the dependence of $|R(x+y/2,x-y/2)|$ from $N$ is dominated by the exponential $e^{-tN^2(\lambda_1+\lambda_2)y^2/2}$, \emph{e.g.} maximally entangled states \cite{Benatti2012-2}, atomic coherent states \cite{Daoud2012}, and the ground state of a double well potential with intra-well interactions \cite{Buonsante2012}.

\section{Conclusions}

In this paper, we have stated the definition of entanglement for identical particles, and identified two classes of states, block-diagonal and non-block-diagonal ones. The latter are dense in the state space. The Hilbert space of pure states, with support on one diagonal block, is isomorphic to that of distinguishable particles; a similar isomorphism does not exist for the whole Hilbert space, when the total number of particles is conserved \cite{Benatti2012,Benatti2012-2}. We have proved that the entanglement of non-block-diagonal states is generated by tunneling, and cannot be erased in finite time for a large class of dissipations, contrary to entanglement of block-diagonal states and distinguishable particles. This has been concretely shown for particle losses and dephasing, where the decay of entanglement turns from exponential into algebraic as the particle number increases. Such robustness is of general interest in physical systems which realize entanglement of identical particles, \emph{e.g.} ultracold gases.

\textbf{Acknowledgments} I acknowledge funding by DFG and by Evaluierter Fonds der Albert-Ludwigs-Universit\"at Freiburg.

\bibliographystyle{unsrt}

\bibliography{ref-ent_dyn_id}

\end{document}